%% file: DyNotears_Inference.tex
\documentclass[conference]{IEEEtran}
\IEEEoverridecommandlockouts

\def\BibTeX{{\rm B\kern-.05em{\sc i\kern-.025em b}\kern-.08em
    T\kern-.1667em\lower.7ex\hbox{E}\kern-.125emX}}
\usepackage{balance}
\usepackage{resizegather}
\usepackage{algorithm}
\usepackage{array}
\usepackage[caption=false,font=normalsize,labelfont=sf,textfont=sf]{subfig}
\usepackage{algpseudocode}
\usepackage{textcomp}
\usepackage{csquotes}
\usepackage{stfloats}
\usepackage{tabstackengine}
\usepackage{multirow}  
\usepackage{booktabs}  
\usepackage{threeparttable}  
\usepackage{url}
\usepackage{verbatim}
\usepackage{graphicx}
\usepackage{cite}
\usepackage{amssymb,amsmath,amsthm,enumitem}

\usepackage[cmintegrals]{newtxmath}
\usepackage{graphicx}
\usepackage{subfig}
\usepackage[export]{adjustbox} 
\begin{document}

\title{Dynamic Structural Causal Models for Delay Prediction}
\thispagestyle{empty}
\author{\IEEEauthorblockN{Niloofar Mehrnia, Wilson Ayyanthole Nelson, and James Gross
}\\
\IEEEauthorblockA{Department of Electrical Engineering and Computer Science, KTH Royal Institute of Technology, Sweden}\\
e-mails:\{nilome,wilsonan,jamesgr\}@kth.se

\thanks{This work has been funded by Vinnova (Sweden’s Innovation Agency) under Grant \#2024-02438.
Niloofar Mehrnia also acknowledges the support of  Vinnova (Sweden’s Innovation Agency) under Grant \#2025-01333.}
}

\maketitle
\thispagestyle{empty}
\begin{abstract}
Dynamic network key performance indicators (KPIs) such as end-to-end (E2E) delay are shaped by both contemporaneous interactions and temporal persistence. While standard predictors can forecast KPIs under a fixed regime, proactive network management requires reasoning about the effects of explicit actions (interventions) and, in time series, executing a model forward over multiple horizons. We study dynamic structural causal models (DSCMs) for E2E delay in a 5G measurement campaign, where payload and background traffic are intentionally manipulated and thus treated as do-interventions. We learn DSCM scaffolds using four time-series causal discovery methods representing score-based and constraint-based paradigms and common lag-based baselines. Using gradient-boosted decision trees, we evaluate true multi-step DSCM rollout, which recursively simulates the full system state, and a direct target-only ablation that predicts the KPI without recursive simulation. Across two held-out intervention scenarios, including high payload and 10\,Mbps background traffic, we find that the inferred target parent/lag sets differ substantially across methods and that rollout behavior is strongly scenario-dependent. In particular, PCMCI+ achieves the best short-horizon rollout accuracy under the payload intervention (RMSE 2.97~ms) but exhibits pronounced horizon-dependent degradation under the background-traffic intervention (root mean squared error (RMSE) $\approx$4.75~ms for long horizon), yielding a large positive rollout tax of up to $\approx$0.78~ms relative to direct prediction; DyNoTEARS and VAR-based DSCMs remain comparatively stable under rollout. These results highlight that evaluating causal scaffolds only via direct prediction can be misleading, and therefore, executability under intervention and long-horizon error propagation are critical for DSCM-based KPI reasoning.
\end{abstract}


\input{Introduction}
\input{SystemModel}
\input{Method}
\input{NumericalResults}
\input{Conclusions}

\balance 

\ifCLASSOPTIONcaptionsoff
  \newpage
\fi
\bibliographystyle{ieeetr}
\bibliography{DyNotears_Inference.bib}

\end{document}

%% file: Introduction.tex
\section{Introduction}
\label{sec:introduction}

The evolution toward 6G networks is widely recognized as a shift toward intelligent, self-optimizing wireless systems, where real-time adaptation and proactive management are fundamental requirements~\cite{AInetwork_01,causal6G_01}. Achieving stringent service-level agreements (SLAs) for latency-, reliability-, and throughput-sensitive applications demands not only accurate prediction of key performance indicators (KPIs), but also decision-relevant inference regarding the impact of network actions and operational regimes~\cite{causality_pearl,inference_01}. In particular, operational control functions such as admission control raise questions at higher levels of Pearl’s causal hierarchy\cite{causality_pearl}: interventions like modifying payload size or introducing background traffic can alter queueing and scheduling dynamics, requiring the controller to estimate how such actions affect the post-intervention latency distribution, rather than simply extrapolating from the pre-intervention regime~\cite{causality_pearl,causality_dibbendu,inference_01}. Correlation-driven models, while often accurate within the training distribution, do not, by construction, distinguish between variables that merely co-vary and those that are genuine causal drivers of a KPI. As a result, associations learned by such models may fail to remain reliable when the operating regime changes, which is precisely the challenge encountered in proactive network control~\cite{causal6G_01,inference_01}.

Traditional approaches to KPI prediction in wireless networks include statistical time-series models such as ARIMA~\cite{kpi_arima_01}, as well as supervised machine learning techniques, such as linear regression, support vector machines (SVMs), and random forests, that are capable of capturing nonlinear relationships between features and KPIs~\cite{kpi_ml_01,kpi_ml_02,kpi_ml_03}. More recently, deep sequence models, particularly deep recurrent architectures like deep recurrent neural networks (DRNNs) and long short-term memory (LSTM) models, have demonstrated strong empirical performance in forecasting nonstationary KPI traces~\cite{kpi_dnn_01,kpi_dnn_02}. However, these methods are primarily optimized for predictive accuracy under the observed data-generating distribution and typically do not provide a principled framework for ``what-if'' (interventional) queries~\cite{causality_pearl,inference_01,causality_dibbendu}. This limitation is particularly significant in autonomous networks, where the objective is often not merely to forecast a KPI under the current regime, but to select actions (e.g., admit or deny traffic, adjust configuration) that causally achieve a target KPI or reduce the probability of SLA violations under uncertainty~\cite{causality_dibbendu,causal6G_01,causal6G_02}.

A dynamic structural causal model (DSCM) offers a framework for such scenarios by integrating (i) a contemporaneous directed acyclic graph (DAG) that captures within-step causal dependencies, and (ii) lagged edges that represent cross-time influences~\cite{causality_timeseries}. To perform multi-step forecasting or interventional reasoning with a DSCM, variables must be predicted in topological order at each future step, with the predicted system state recursively rolled forward across multiple horizons~\cite{causality_timeseries}. This execution paradigm introduces a critical practical challenge: even if the one-step mechanisms are highly accurate, errors can accumulate and propagate through time and across intermediate nodes during recursive rollout, resulting in horizon-dependent degradation, a phenomenon that remains hidden if causal discovery is used solely as a feature-selection tool~\cite{causality_timeseries}.

In this work, we empirically investigate the impact of DSCM execution on packet-level end-to-end (E2E) latency under controlled regime changes. We treat payload size and background traffic rate as experimenter-controlled configuration variables, interpreting changes in these parameters as do-interventions~\cite{causality_pearl,inference_01}. Our study compares four state of the art time-series causal discovery methods, Dynamic Non-combinatorial Optimization via Trace Exponential and Augmented lagRangian for Structure learning (DyNoTEARS), Peter-Clark Momentary Conditional Independence (PCMCI+), Vector Autoregressive Granger (VAR-Granger), and VAR-Linear Non-Gaussian Acyclic Model (VAR-LiNGAM), by fitting a DSCM mechanism class and evaluating (a) \emph{learned DSCM rollout}, which recursively predicts the full system state, versus (b) a \emph{direct} target-only predictor that leverages the discovered target parents but does not recursively simulate the entire system. This approach enables a direct quantification of the ``rollout tax'' attributable to error propagation during DSCM execution~\cite{causality_timeseries}.
The main contributions of the paper are as follows:
\begin{itemize}
    \item We develop and implement a DSCM execution and evaluation framework for packet-level E2E delay under controlled regime changes in payload and background traffic, explicitly treating these as do-interventions on exogenous configuration variables~\cite{causality_pearl,inference_01}.
    \item We systematically compare DyNoTEARS, PCMCI+, VAR-Granger, and VAR-LiNGAM in terms of (i) the inferred parent and lag structure for the E2E delay node, and (ii) the downstream performance of the resulting executable DSCM, using a unified mechanism estimation and rollout framework.
    \item We introduce and report an error-propagation metric, the \emph{rollout tax}, defined as rollout RMSE minus direct RMSE, across multiple horizons, and demonstrate that this tax is both scenario- and method-dependent, revealing failure modes that are not apparent under direct prediction.
\end{itemize}

The remainder of this paper is structured as follows. Section~\ref{sec:system_model} outlines the system model assumptions and the problem formulation. In Section~\ref{sec:method}, we detail the DSCM learning algorithms. Section~\ref{sec:numerical_results} discusses the measurement data, model assumptions, and provides a comprehensive evaluation. Finally, Section~\ref{sec:conclusions} concludes the paper.

%% file: SystemModel.tex
\section{System Model and Problem Formulation}
\label{sec:system_model}

We consider a packet-based 5G radio access scenario in which a user equipment (UE) transmits traffic periodically, and the network provides per-packet measurements that capture both configuration variables and performance indicators. Each transmitted packet is treated as a single time step, corresponding to the packet transmission periodicity. The primary objective is to model the E2E delay KPI and to enable \emph{interventional} queries that assess how this KPI responds to controlled changes in operating conditions, in addition to forecasting under regime shifts~\cite{causality_pearl,inference_01}.

\subsection{System Assumptions}
\label{sec:observed_variables}
At each packet index $t$, we observe a multivariate vector
\begin{equation}
X_t = \big[X_t^{(1)},\ldots,X_t^{(d)}\big] \in \mathbb{R}^d,
\end{equation}
where the target KPI is defined as
\begin{equation}
Y_t \equiv \text{E2E delay at packet } t \quad (\texttt{End to End Delay}).
\end{equation}
In our system, $X_t$ comprises a selected subset of packet-level and RAN-level variables, including packet length (\texttt{Packet Length}), link adaptation parameters (\texttt{mcs}), retransmission and attempt indicators (\texttt{No of RLC attempts}), and protocol delay components (\texttt{segmentation delay}), as well as the target KPI itself.
In addition to per-packet measurements, the dataset includes predetermined \emph{run-level regimes}. Specifically, payload size (represented by \texttt{Packet Length}) and background traffic rate (represented by \texttt{bg\_rate\_mbps}) are externally configured to systematically investigate their impact on E2E delay. Consequently, we treat these regime variables as \emph{exogenous intervention targets} within the causal model~\cite{causality_pearl,inference_01}. In practice, this means that during structure learning and post-processing, we explicitly disallow any incoming causal edges to these regime nodes.

\subsection{Dynamic Structural Causal Model (DSCM)}
\label{sec:dscm}
To capture temporal dependencies and enable executable intervention reasoning, we employ a DSCM~\cite{causality_timeseries}, which is characterized by a time-indexed causal graph,
\begin{equation}
G_T = \big(V,\{E_\ell\}_{\ell=0}^{p}\big),
\end{equation}
where $V$ denotes the set of variables, $E_0$ represents contemporaneous (same-step) directed edges, and $E_\ell$ for $\ell \geq 1$ corresponds to lagged edges that capture influences from time $t-\ell$ to time $t$. The contemporaneous graph $E_0$ is assumed to be acyclic (DAG), which allows a topological ordering of variables for causal evaluation~\cite{causality_timeseries}.
For each variable $x_k \in V$, let $\mathrm{PA}_k$ denote its contemporaneous parents and $\mathrm{LPA}_k$ denote its lagged parents up to lag $p$. The DSCM defines a structural mechanism for each variable as
\begin{equation}
\label{eqn:dscm}
x_{k,t} = g_k\!\left(\mathrm{PA}^{0}_{k,t}, \mathrm{PA}^{1:p}_{k,t}, \varepsilon_{k,t}\right),
\end{equation}
where $\mathrm{PA}^0_{k,t}$ and $\mathrm{PA}^{1:p}_{k,t}$ are the contemporaneous (same-step) parents and the lagged parents (up to $p$ lags) of node $k$ at time $t$, respectively, $g_k(\cdot)$ is an unknown function learned from data, implemented in this work using gradient-boosted regression trees, and $\varepsilon_{k,t}$ is the error term. Incorporating lagged parents reduces the need to explicitly model latent disturbances during multi-step prediction, allowing the system to be executed by recursively predicting the full state forward in time~\cite{causality_timeseries}.

\subsection{Causal Interventions}
\label{sec:interventions}
Employing standard do-intervention, including $do(\texttt{Packet Length}=\ell)$ or $do(\texttt{bg\_rate\_mbps}=b)$, indicates that the corresponding node is externally fixed to the specified value, thereby severing any incoming causal links to that node~\cite{causality_pearl,inference_01}. Under such interventions, the DSCM is executed by evaluating~\eqref{eqn:dscm} for all endogenous nodes in topological order at each time step, and recursively rolling the predicted state forward across multiple horizons~\cite{inference_01}.

\subsection{Problem Formulation}
\label{sec:problem_formulation}

Given training runs collected under baseline regimes, our objectives include: a) DSCM learning, b) executable multi-step forecasting under regime shift, and c) ablation against direct prediction. We perform DSCM learning to learn a causal scaffold $G_T$, comprising both instantaneous and lagged edges, and to fit structural mechanisms $\{g_k\}$ for all endogenous nodes. Then, for executable multi-step forecasting under regime shift, we evaluate, on test runs corresponding to interventions (e.g., high payload and high background traffic), the $H$-step prediction accuracy of \emph{learned DSCM rollout}, which entails recursively executing the learned DSCM for horizons $h=1,\dots,H$~\cite{inference_01}. Afterwards, for ablation against direct prediction, we compare DSCM rollout with a \emph{direct} target-only predictor that uses the same discovered target parents and lags, but predicts $Y_{t+h}$ directly from the observed history at time $t$ without recursive simulation. This comparison isolates the effect of error accumulation introduced by recursive execution.

To quantify the extent of horizon-dependent degradation caused by recursive execution, we define the \emph{rollout tax} at horizon $h$ as
\begin{equation}
\mathrm{tax}(h) \triangleq \mathrm{RMSE}_{\mathrm{rollout}}(h) - \mathrm{RMSE}_{\mathrm{direct}}(h).
\end{equation}
A positive rollout tax indicates that recursive DSCM rollout yields higher prediction error than direct prediction at that horizon, reflecting the accumulation and propagation of errors through intermediate predictions during multi-step execution.



%% file: Method.tex
\section{Learning and Executing Dynamic SCMs}
\label{sec:method}
This section outlines the construction of a causal model for packet-level KPI reasoning. We begin by clarifying the distinction between static SCMs and DSCMs. Next, we describe the causal discovery methods considered in this work, including score-based, constraint-based, and vector autoregressive (VAR)-based approaches. Finally, we present the unified downstream mechanism learning and evaluation protocol applied to all discovered causal scaffolds.

\subsection{From SCM to DSCM}
\label{sec:scm_vs_dscm}

A static structural causal model (SCM) represents each variable as a structural assignment of the form $x_k = g_k(\mathrm{PA}_k, \varepsilon_k)$, where $\mathrm{PA}_k$ denotes the (static) set of parent variables and $\varepsilon_k$ is an exogenous noise term~\cite{causality_pearl,inference_01}. While SCMs provide a principled framework for modeling interventions via the do-operator, they do not inherently capture temporal persistence: repeating the same intervention does not propagate state information across time.

In contrast, packet-level KPIs in wireless networks exhibit pronounced temporal dependencies. To address this, we adopt a DSCM, in which (i) contemporaneous causal relationships within a time step are encoded by a directed acyclic graph (DAG) $E_0$, and (ii) temporal dependencies are represented by lagged edges $E_\ell$ for $\ell \geq 1$, up to a maximum lag $p$. For node $k$ at time $t$, the DSCM mechanism is expressed in (\ref{eqn:dscm}).

A key implication of the DSCM framework is that multi-step prediction and interventional forecasting require explicit \emph{execution} of the model: at each time step, variables are predicted in the topological order defined by $E_0$, conditioning on lagged parents from the previous $p$ steps, and the predicted state is then recursively propagated forward. This recursive execution can amplify modeling errors across prediction horizons, a phenomenon known as error propagation, which motivates evaluating DSCMs not only as feature selectors but also as an executable framework.

\subsection{Causal structure learning methods}
\label{sec:structure_learning}

We compare four approaches for learning the DSCM scaffold $G=(V,W,A_{l}) \, , l\in \{1:p\}$ from multivariate time series.

\subsubsection{DyNoTEARS}
This \emph{score-based} framework models the data as a linear structural VAR with instantaneous and lagged effects,
\begin{equation}
X_t = X_t W + \sum_{\ell=1}^p X_{t-\ell} A_\ell + Z_t,
\end{equation}
and learns the weight matrix of contemporaneous, $W$, and lagged, $\{A_\ell\}$, by minimizing a regularized least-squares objective with a smooth acyclicity constraint on the DAG \cite{dynotears}. The $\ell_1$ penalties encourage sparsity in both the contemporaneous matrix and the lagged matrix. 

\subsubsection{PCMCI+}
It is a \emph{constraint-based} method for time-series causal discovery that identifies lagged dependencies by performing conditional independence tests, while controlling the false positive rate through a staged selection procedure~\cite{pcmci}. In our implementation, PCMCI+ is primarily used to determine the set of lagged edges $E_{\ell}$ for $\ell=1,\ldots,p$. Contemporaneous edges are treated as 0, as their orientation can be ambiguous at finite sampling resolutions and may introduce cycles into the graph.

\subsubsection{VAR-Granger}
This lag-based approach selects lagged parents for each variable by conducting hypothesis tests on the coefficients of a VAR model~\cite{vergranger}. Specifically, for each variable $x_k$, lagged parents are identified based on the statistical significance of the corresponding lagged coefficients in the VAR model, resulting in a directed scaffold that includes only lagged edges $E_{\ell}$ for $\ell=1,\ldots,p$. We use this as a predictive causal baseline: while Granger-causal edges indicate incremental predictability, they do not necessarily imply interventional causality in the presence of unobserved confounding.

\subsubsection{VAR-LiNGAM}
This framework is a hybrid approach that first fits a VAR model to identify lagged dependencies, and then applies a LiNGAM-style method to the residuals in order to orient contemporaneous (instantaneous) relations under the assumption of non-Gaussianity~\cite{varlingam}. This procedure yields a candidate DAG for the contemporaneous edges $E_0$, in addition to the lagged edges $E_{\ell}$ for $\ell=1,\ldots,p$ inherited from the VAR component. We include VAR-LiNGAM in our comparison to assess whether the additional orientation of instantaneous edges leads to differences in downstream executable behavior.

\subsection{Causal Inference}
\label{sec:mechanism_learning}

Given a discovered scaffold $(E_0, E_{1:p})$, we fit a structural mechanism $g_k(\cdot)$ for each endogenous node $x_k$ using supervised regression on its corresponding parent sets. Specifically, for each node $x_k$, the parent sets consist of contemporaneous parents $\mathrm{PA}^0_{k,t}$ (from $E_0$) and lagged parents $\mathrm{PA}^{1:p}_{k,t}$ (from $E_{1:p}$). We employ gradient-boosted regression trees (XGBoost) as a flexible function class for learning each $g_k(\cdot)$.

For contemporaneous parents, we adopt a teacher-forcing approach during training: when predicting $x_{k,t}$, we condition on the observed values of already-ordered same-step parents, following a topological ordering of $E_0$. At execution time, these contemporaneous parent values are generated sequentially according to the same topological order, ensuring consistency between training and inference.

Given the last $p$ observed time steps, we predict the next state $X_{t+1}$ by traversing the contemporaneous edge set $E_0$ in topological order and evaluating each structural mechanism $g_k$ using both contemporaneous and lagged parents. This process is then recursively repeated to obtain $X_{t+h}$ for prediction horizons $h=1,\dots,H$. The forecast for the end-to-end (E2E) delay is extracted from the corresponding component of $X_{t+h}$.

To isolate the effect of recursive execution, we also evaluate a direct predictor that uses a horizon-specific XGBoost regressor that predicts $Y_{t+h}$ directly from the observed history at time $t$ using only the discovered parents (and lags) of the target $Y$, without recursively simulating the full state. Comparing direct prediction to true DSCM rollout provides an empirical measure of the error accumulation attributable to recursive execution in the DSCM.


%% file: NumericalResults.tex
\section{Numerical Results}
\label{sec:numerical_results}

We evaluate DSCMs for packet-level E2E delay prediction under controlled regime shifts. Our primary aim is \emph{not} to assert that causal modeling universally enhances pure forecasting accuracy. Instead, we focus on (i) examining how different time-series causal discovery methods yield distinct DSCM scaffolds, and (ii) analyzing the behavior of these scaffolds when executed as \emph{true multi-step DSCM rollouts}, where recursive execution may introduce horizon-dependent error propagation. Additionally, we include a direct (target-only) ablation using XGBoost to disentangle the impact of recursive rollout from the influence of the parent sets.

\subsection{Measurement Data and Intervention Scenarios}
\label{sec:data}

To enable a comprehensive evaluation of our E2E latency prediction approach, we instrument a 5G network testbed, named ExPECA \cite{expeca}, using the End-to-End Delay Analytics Framework (EDAF)~\cite{edaf}. The testbed is based on an OpenAirInterface (OAI) 5G New Radio (NR) deployment with software-defined radio (SDR) gNB and UE nodes, precise time synchronization via PTP, and multi-layer packet timestamping (LatSeq). 
We train the model on 1,000 measurements collected per run, under baseline regimes with periodic traffic at 50\,ms intervals and no background traffic, spanning payload sizes of 20/50/100/200/500/1000/1500\,B. To broaden regime coverage, the training dataset also includes runs with background traffic generated by additional UEs (100\,bps and 2.5/5/7.5\,Mbps). Each packet corresponds to one time step under the 50\,ms periodic workload. During the measurement campaign, payload and background traffic are explicitly set by the experimenter to probe their impact on E2E delay. Consequently, cross-run changes in these variables are treated as do-interventions on exogenous regime nodes~\cite{causality_pearl,inference_01}.

In the results below, we report two held-out test scenarios, each evaluated on 1{,}000 packet samples:
\begin{itemize}
    \item Scenario 1 (background-traffic intervention): background traffic fixed at 10\,Mbps.
    \item Scenario 2 (payload intervention): payload fixed at 2\,KB.
\end{itemize}
The DSCM includes the variables \texttt{Packet Length}, \texttt{bg\_rate\_mbps}, \texttt{tx\_gain\_db}, \texttt{mcs}, \texttt{No of RLC attempts}, \texttt{segmentation delay}, and \texttt{End to End Delay}. We treat \texttt{Packet Length} and \texttt{bg\_rate\_mbps} as exogenous intervention targets (i.e., no incoming edges) to align the learned graph with do-semantics~\cite{causality_pearl,inference_01}. Finally, \texttt{segmentation delay} is a link-delay subcomponent that becomes prominent when the packet size exceeds the transport block size.

\subsection{Model Parameters}
\label{sec:eval_protocol}

DyNoTEARS is configured with $\lambda_W=\lambda_A=10^{-4}$ and an edge-weight threshold of $0.1$, so that estimated instantaneous and lagged coefficients with magnitude below $0.1$ are pruned from the scaffold~\cite{dynotears}. For PCMCI+, we use a significance threshold of $\alpha=0.05$ for the conditional independence tests~\cite{inference_01}. For VAR-Granger, lagged edges are selected via hypothesis tests on VAR coefficients (up to lag $p=5$), yielding a lag-only scaffold; for VAR-LiNGAM, we additionally obtain a candidate instantaneous DAG $E_0$ by orienting contemporaneous relations in the VAR residuals under non-Gaussianity assumptions~\cite{inference_01}.
Unless stated otherwise, we use $N_{\mathrm{train}}=1000$ samples per training run and $N_{\mathrm{test}}=1000$ samples per test run. We evaluate multi-step prediction over horizons $h=1,\ldots,H$ with $H=20$. The XGBoost regressor used here as a direct predictor utilizes 2,500 trees with a moderate depth of 6 and a low learning rate of 0.03 to ensure gradual, stable learning.

\subsection{Discussions}
\label{sec:target_parents}

\begin{figure}[t]
    \centering
    \captionsetup[subfigure]{labelformat=empty}
     \begin{center}
        \subfloat[(a)]{%
            \label{fig:barplot}
            \includegraphics[width=.9\columnwidth]{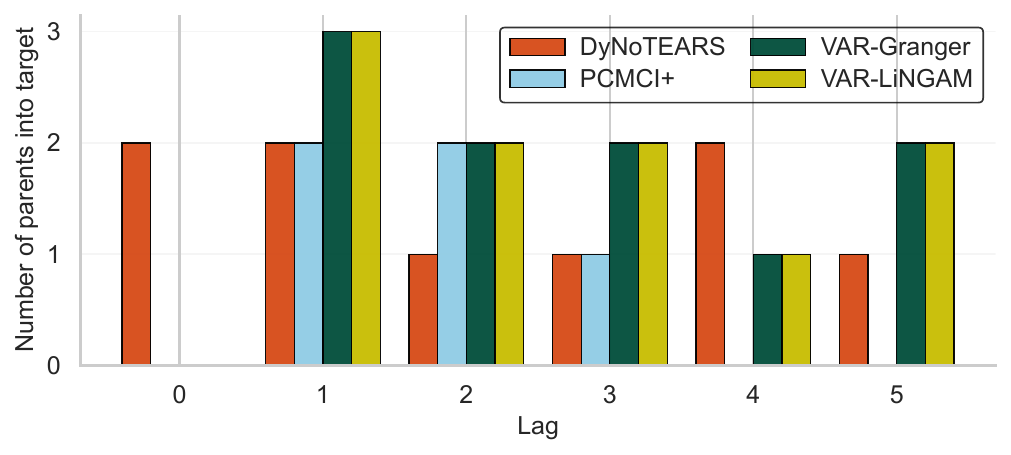}
        }\\
        \subfloat[(b)]{%
            \label{fig:matrix}
            \includegraphics[width=.95\columnwidth,height=4cm]{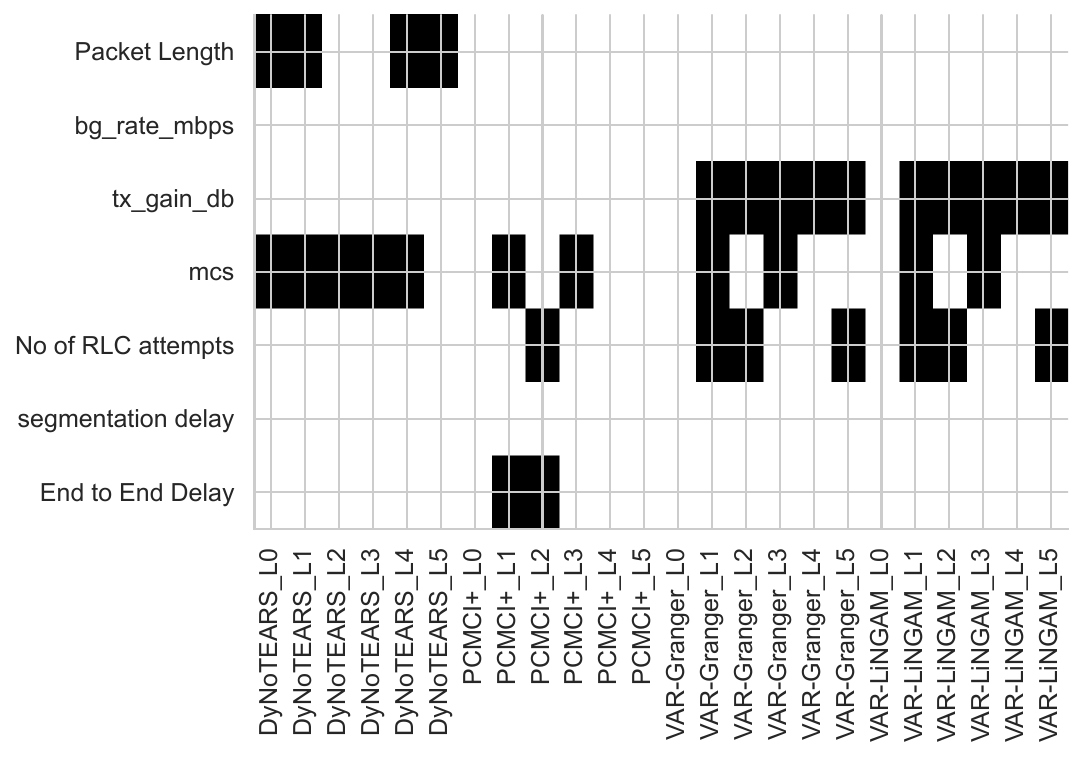}
        }\\
    \end{center}
    \vspace{-2mm} 
    \caption{Causal parents (with lags) into \texttt{End to End Delay} discovered by DyNoTEARS, PCMCI+, VAR-Granger, and VAR-LiNGAM. lag 0 means instantaneous): a) number of parents, and b) exact features.
    }
    \label{fig:target_parents}
\end{figure}

Fig.~\ref{fig:target_parents} illustrates causal parents with lags into \texttt{End to End Delay} discovered by DSCM methods, DyNoTEARS, PCMCI+, VAR-Granger, and VAR-LiNGAM.
The discovered parent/lag sets for the target node (\texttt{End to End Delay}) differ substantially across methods, reflecting different implied ``memory'' and mediating structure. For example, DyNoTEARS selects \texttt{Packet Length} and \texttt{mcs} at multiple lags as parents of E2E delay, while PCMCI+ includes explicit E2E self-lags (e.g., lag 1 and lag 2) as well as \texttt{mcs} and \texttt{No of RLC attempts} at selected lags. VAR-based methods include multi-lag dependencies on \texttt{tx\_gain\_db}, \texttt{No of RLC attempts}, and \texttt{mcs}. These structural differences may affect both (i) short-horizon predictive performance and (ii) stability under recursive rollout, where intermediate prediction errors may compound over time. For instance, PCMCI+ includes E2E self-lags, while DyNoTEARS emphasizes payload/MCS lags, which influence rollout stability.

\begin{figure}[t]
    \centering
    \captionsetup[subfigure]{labelformat=empty}
     \begin{center}
        \subfloat[(a)]{%
            \label{fig:scatterplot_17}
            \includegraphics[width=.8\columnwidth]{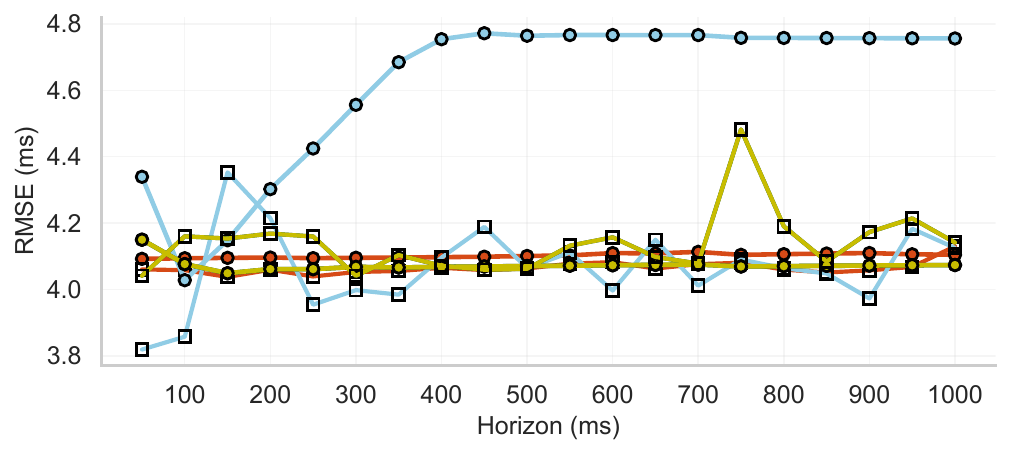}
        }\\ \vspace{-0.2cm}
        \subfloat[(b)]{%
            \label{fig:scatterplot_8}
            \includegraphics[width=.8\columnwidth]{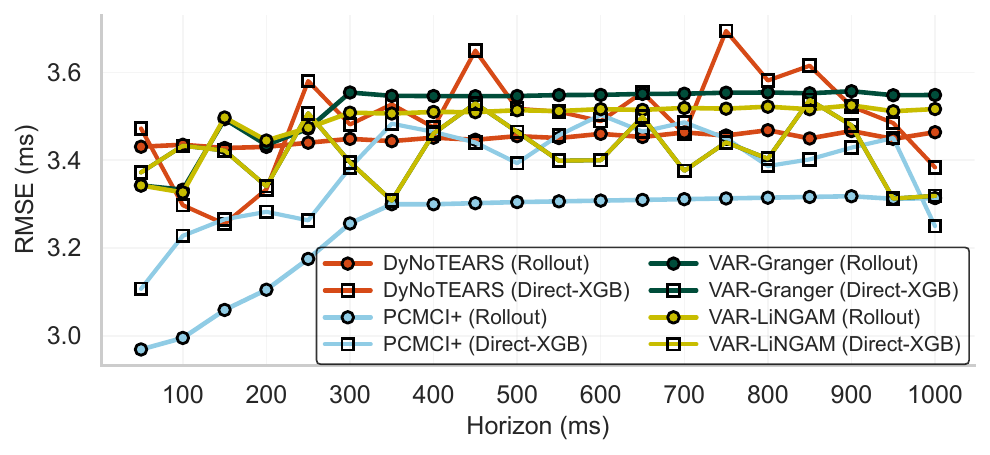}
        }\\
    \end{center}
    \vspace{-2mm} 
    \caption{RMSE vs. horizon evaluated on: a) 10\, Mbps background traffic, and b) payload of 2000B per 50~msec. Legend in (b) corresponds to both figures. 
    }
    \label{fig:rmse_horizon}
\end{figure}

Fig.~\ref{fig:rmse_horizon} presents the RMSE as a function of prediction horizon for the true DSCM rollout and the direct ablation, evaluated separately under two causal-intervention conditions (figure legend is shown in Fig.~\ref{fig:rmse_horizon}(b)). In the 10\,Mbps background-traffic setting shown in Fig.~\ref{fig:rmse_horizon}(a), the PCMCI+ rollout shows pronounced degradation with increasing horizon, rising from approximately 4.03-4.34~ms at short horizons to about 4.75-4.77~ms for horizons beyond roughly eight steps. In contrast, the DyNoTEARS and VAR-based rollouts remain comparatively stable, maintaining RMSE values around 4.07-4.11~ms across horizons. This demonstrates that the choice of structural scaffold significantly influence long-horizon rollout stability under a regime shift.
In the high-payload test scenario shown in Fig.~\ref{fig:rmse_horizon}(b), PCMCI+ achieves the lowest short-horizon RMSE (e.g., 2.97~ms at $h = 1$) but increases gradually with horizon, reaching approximately 3.31~ms by $h = 20$. DyNoTEARS remains relatively stable across horizons (around 3.43-3.47~ms). VAR-based rollouts also remain stable but generally exhibit higher errors than PCMCI+ in this regime.

\begin{figure}[t]
    \centering
    \captionsetup[subfigure]{labelformat=empty}
     \begin{center}
        \subfloat[(a)]{%
            \label{fig:scatterplot_17}
            \includegraphics[width=.8\columnwidth]{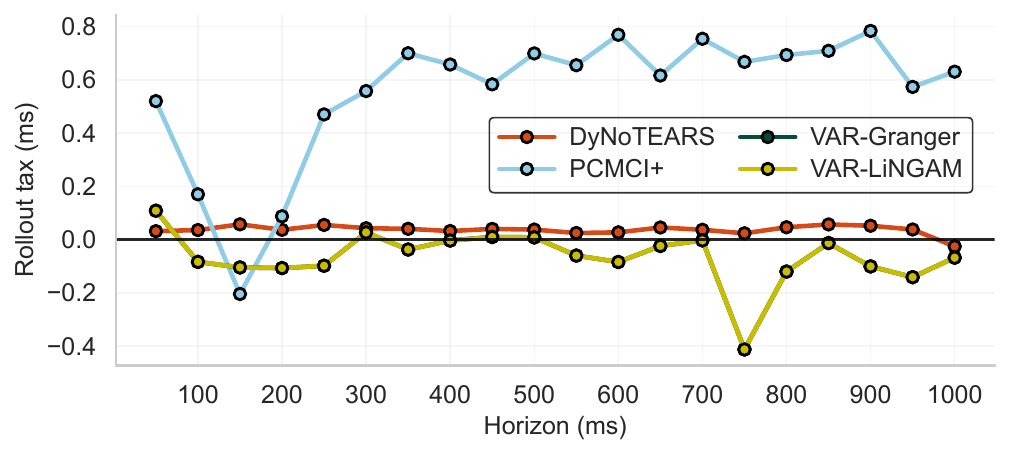}
        }\\ \vspace{-0.2cm}
        \subfloat[(b)]{%
            \label{fig:scatterplot_8}
            \includegraphics[width=.8\columnwidth]{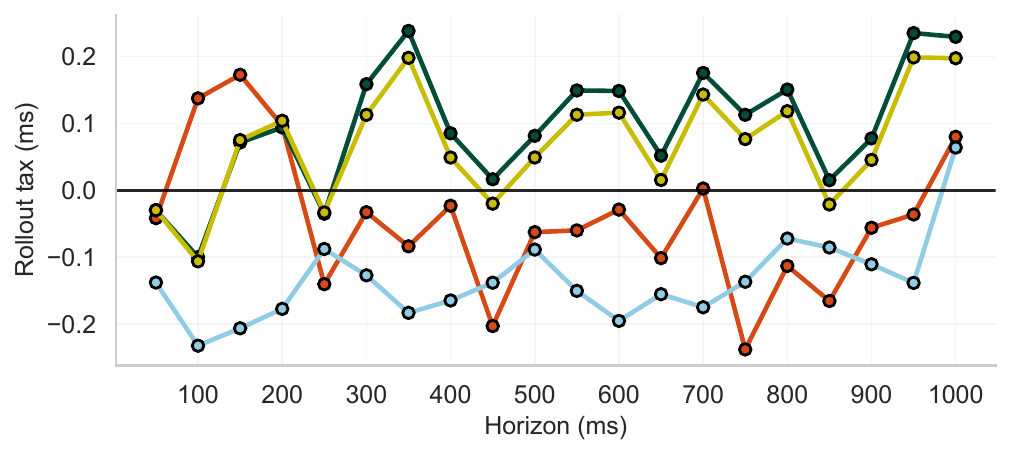}
        }\\
    \end{center}
    \vspace{-2mm} 
    \caption{Rollout tax vs. horizon under: a) 10\,Mbps background traffic, and b) payload of 2000B per 50~ms. Legend in (a) corresponds to both figures. 
    }
    \label{fig:tax}
\end{figure}

Fig.~\ref{fig:tax} presents the error-propagation analysis using rollout tax under two intervention scenarios. Consistent with the previous results, the figure shows that the rollout tax varies across both scenarios and modeling approaches. In the background-traffic intervention shown in Fig.~\ref{fig:tax}(a), PCMCI+ exhibits a substantial positive tax across most horizons (approximately 0.5-0.78~ms), indicating significant error accumulation during recursive rollout. This can be due to conditioning sets/CI tests behaving poorly under regime shift, or because PCMCI+ includes E2E self-lags, which can amplify bias in rollouts. DyNoTEARS maintains a near-zero tax over most horizons, reflecting stable execution under this intervention. VAR-based methods display tax values ranging from small to moderate, depending on the horizon.
In the high-payload intervention scenario shown in Fig.~\ref{fig:tax}(b), PCMCI+ frequently yields a negative tax, meaning that rollout outperforms direct prediction, suggesting that a single learned one-step executable model can regularize predictions more effectively than horizon-specific direct regressors in this regime.


These results support two practical insights. First, different time-series causal-discovery methods can infer noticeably different causal scaffolds for the same KPI, including variations in the degree of autoregressive dependence and the role of mediating variables. Second, assessing a causal scaffold solely through direct prediction can obscure important failure modes: when deployed as a full DSCM rollout, required for long-horizon forecasting and interventional trajectory simulation, some scaffolds incur substantial error-propagation under regime shift, whereas others remain stable. This highlights the importance of reporting executable-rollout performance and rollout tax as core evaluation metrics for DSCM-based KPI reasoning, particularly in intervention-driven network-management settings. Table~\ref{tab:comp} summarizes the observed advantages and limitations of each DSCM learning framework.

\begin{table}[]
    \centering
    \scriptsize
    \caption{Table of Pros and Cons of each DSCM learning framework}
    \begin{tabular}{p{1.5cm}||p{3cm}||p{3cm}}
    \hline
        Method & Pros & Cons \\
        \hline
        DyNoTEARS & Robust across scenarios and regime shift; stable RMSE; & Not always best at short horizons \\
        \hline
         PCMCI+ & Best short-horizon RMSE; strong lag detection & Error growth at long horizons; sensitive to regime shift \\
         \hline
         VAR-Granger & Stable long-horizon RMSE; competitive in low-payload & May miss contemporaneous effects \\
         \hline
         VAR-LiNGAM & Same as VAR-Granger in this setting & No added value over VAR-Granger\\
         \hline
        
    \end{tabular}
    
    \label{tab:comp}
\end{table}

We considered one additional metric to further evaluate the performance of different DSCM (Rollout) frameworks: Bias quantifies whether a model systematically over- or underestimates the true \texttt{End to End delay}. A positive bias means the model tends to overpredict, while a negative bias indicates underprediction. In the high background traffic intervention scenario, where the distribution of \texttt{End to End delay} shifts more significantly, PCMCI+ stands out with a pronounced positive bias, with a mean value of 2.2~ms, indicating a consistent tendency to overestimate delay. In contrast, all other methods show lower bias, typically 0.25-0.65~ms, suggesting that PCMCI+ is susceptible to error accumulation in recursive prediction, leading to persistent overestimation. 

%% file: Conclusions.tex
\section{Conclusions}
\label{sec:conclusions}

We investigated dynamic structural causal models (DSCMs) for packet-level end-to-end delay under controlled regime shifts, comparing DyNoTEARS, PCMCI+, VAR-Granger, and VAR-LiNGAM as causal-scaffold learners while holding the downstream mechanism-learning stage fixed. The inferred target parent and lag sets differ substantially across methods, revealing distinct forms of ``target memory'' and mediator structure. For instance, PCMCI+ includes explicit self-lags for E2E delay, whereas DyNoTEARS emphasizes multi-lag dependencies on \texttt{Packet Length} and \texttt{mcs}. Our results emphasize the importance of \emph{executable rollout}. When it is deployed as genuine multi-step DSCM rollouts, some scaffolds exhibit strong horizon-dependent error propagation under the background-traffic intervention, whereas others remain comparatively stable. This behavior is directly highlighted by the rollout-versus-direct ablation through the rollout-tax metric. Future work will validate counterfactuals using explicit disturbance abduction and evaluate ablations that remove delay-component variables, such as \texttt{segmentation delay}, to assess the model's sensitivity to near-target information.